\documentclass[a4paper,11pt]{llncs}

\usepackage{pgfplots} 
\usepackage{graphicx}
\usepackage{soul}
\usepackage{algorithm}
\usepackage[noend]{algpseudocode}
\usepackage{paralist}
\usepackage{flushend}
\usepackage{authblk}
\usepackage{multirow}
\usepackage{amsmath}
\DeclareMathOperator*{\argmax}{argmax}

\begin{document}

\title{Capture the Bot: Using Adversarial Examples to Improve CAPTCHA Robustness to Bot Attacks}

\author{Dorjan Hitaj\inst{1} \and Briland Hitaj\inst{2} \and
Sushil Jajodia\inst{3}  \and Luigi V. Mancini\inst{1}
}

\institute{Dipartimento di Informatica, Sapienza Universit\`{a} di Roma, Italy\\ 
\email{\{hitaj.d, mancini\}@di.uniroma1.it},\\ 
\and
Computer Science Laboratory, SRI International\\
\email{briland.hitaj@sri.com},\\
\and
Center for Secure Information Systems, George Mason University\\
\email{jajodia@gmu.edu}}

\maketitle

\markboth{Oct~2020}
{Shell \MakeLowercase{\textit{Hitaj et al.}}: Capture the Bot: Using Adversarial Examples to Improve CAPTCHA Robustness to Bot Attacks}

\begin{abstract}

To this date, CAPTCHAs have served as the first line of defense preventing unauthorized access by (malicious) bots to web-based services, while at the same time maintaining a trouble-free experience for human visitors. However, recent work in the literature has provided evidence of sophisticated bots that make use of advancements in machine learning (ML) to easily bypass existing CAPTCHA-based defenses. 
In this work, we take the first step to address this problem. We introduce CAPTURE, a novel CAPTCHA scheme based on adversarial examples. While typically adversarial examples are used to lead an ML model astray, with CAPTURE, we attempt to make a ``good use'' of such mechanisms. Our empirical evaluations show that CAPTURE can produce CAPTCHAs that are easy to solve by humans while at the same time, effectively thwarting ML-based bot solvers.

\end{abstract}

\thispagestyle{plain}
\pagestyle{plain}

\section{Introduction}
\label{sec:introduction}

Completely Automated Public Turing test to tell Computers and Humans Apart (CAPTCHA) is a type of challenge-response test in computing applications which is used to distinguish between humans and machines.

Usually CAPTCHAs are generated by distorting an image that contains text and numbers in such a way that only a human can properly read or understand the content of that image. 
However, current advancements in Machine Learning (ML), particularly in the domain of Deep Learning (DL), pose a significant threat to existing CAPTCHA mechanisms. Deep Neural Networks (DNNs), the core component of deep learning, have even surpassed human capabilities in certain tasks, such as image and speech recognition. Recent research also shows that DNNs are able to bypass existing CAPTCHA-based defenses, opening the doors to automated attacks on virtually any online service that relies on CAPTCHA as a defense mechanism.

To counteract this phenomenon, CAPTCHA designers are trying to make the CAPTCHA task harder to solve by introducing additional security features. However, most of these additional security features end up impacting the user experience as well. 
Some types of CAPTCHAs, such as the Google image-selection CAPTCHA, are becoming increasingly hard to solve even for humans due to the difficulty of recognizing the small details shown in the challenge images. In turn, this additional burden leads to frustration whenever users are presented with a complex CAPTCHA that requires a significant concentration to be solved, just to be able to access the desired online service.

Our paper introduces a novel technique, called CAPTURE - \textbf{CAP}tcha \textbf{T}echnique \textbf{U}niquely \textbf{RE}sistant, based on adversarial machine learning to improve the resilience of image-selection CAPTCHAs against automated solvers, while at the same time maintaining a positive user experience in the CAPTCHA solving process.
Specifically, we investigate the "good use" of unrecognizable images~\cite{unrecognizableImages} and adversarial patch~\cite{adversarialPatch} to design an improved image-selection CAPTCHA. 
\textit{We show that unrecognizable images and adversarial patches can be used to alter images without affecting human-based recognition and user experience, while at the same time leading DNN-based automated solvers to misclassify the images with high confidence. }
Differently from traditional adversarial perturbations~\cite{intriguingSzegedy}, unrecognizable images and adversarial patches are highly transferable between various DNN model architectures~\cite{papernot2016transferability}, which makes these two techniques good building blocks to design a general countermeasure against machine learning based CAPTCHA solvers. 

This paper focuses on DNN-based automatic solvers, since DNN models have proven to be the go-to machine learning technique to handle complex tasks such as image recognition, thus being among the most widespread techniques employed by automatic solvers to solve a CAPTCHA challenge.

We conduct security and usability experiments for our novel CAPTURE technique. Our experiments show that the CAPTCHAs created using CAPTURE are hard to solve by bots, but are easily solved by humans in a short amount of time. 

\section{Background}
\label{sec:background}
\subsection{(Deep) Neural Networks}
Neural networks are a set of algorithms that focus on determining underlying relationships existing in the input data in a relatively similar way to the human brain.
Deep neural networks (DNN) differ from traditional single-layer neural networks by their number of hidden layers. Typically, a DNN consists of two or more hidden layers. Stacking more layers in a DNN enables the later to identify and extract more complex features from the input data. This hierarchy-based structure enables DNNs to process tremendous amounts of high-dimensional data, such as images, with impressive results both in classification and clustering tasks. 
Despite the superhuman capabilities that DNNs have shown on performing certain tasks, they are susceptible to adversarial attacks that can deteriorate their performance significantly. This vulnerability towards adversarial attacks makes entities approach the domain with care even though the benefits of using DNNs seem undoubtedly large.
Even though this property of DNNs seems rather negative, we attempt to make a good use out of it, by leveraging it to introduce security features that make current CAPTCHA schemes harder to solve by automated CAPTCHA solvers that leverage DNNs in the CAPTCHA solving process.

\subsection{Adversarial Examples}
Adversarial examples are maliciously crafted inputs whose purpose is to lead a machine learning model into misclassifying the input. They are intended to be stealthy in a such a way that the perturbation that will be added to a legitimate input instance will not lead a human into misclassifying it.
These malicious inputs were first observed by Szegedy et al.~\cite{intriguingSzegedy}. Since that time, extensive research is done on both directions, introducing new adversarial example generation techniques and new protection techniques against them.
Moreover, extensive work has been done into creating models that are robust to adversarial perturbations, such as adversarial training~\cite{ensembleTramer}.
Adversarial training is a training procedure that makes ML models more robust to adversarial examples, but not completely safe from them. To this date, there is no universal defense against adversarial examples. All the published defenses have been taken down shortly after by new adversarial perturbation generation methods that are stealthier and stronger.

Typically, adversarial examples are generated by exploiting the internals of a machine learning model, which makes them more effective and stealthier. Moreover, Papernot et al.~\cite{papernot2016transferability} observed that some adversarial examples are also transferable, meaning that adversarial examples that affect one ML model often affect another ML model, even if the two models have different architectures or were trained on different training sets, as long as both models are trained to perform the same task. This means that an attacker may train their own substitute model, craft adversarial examples against the substitute, and transfer them to a victim model, with very little information about the victim. 

Inspired by the transferability property, Shi et al.~\cite{adversarialCaptcha} present a methodology to harden CAPTCHASs by using adversarial perturbations in order to make ML-based Bots unable to solve the presented CAPTCHA challenge. They present a new adversarial perturbation technique that adds perturbations in the frequency domain rather than in the space domain. According to their analysis, the perturbations added in the space domain are frail to image preprocessing and are considered as local change to images while the perturbations added in the frequency domain are a global change, thus more difficult to remove. Even though this technique is resilient to various perturbation removal methods, it still suffers from the inconsistency of the transferability property among various ML models. To this end, we aim at employing methods that allow us to create highly transferrable instances in order to provide a novel CAPTCHA scheme that is more resilient to Bots. Moreover the methods introduced in this paper, enable us to also improve the user experience in the CAPTCHA solving process.

\subsection{Unrecognizable Images}

Nguyen et al.~\cite{unrecognizableImages} show a related result to adversarial perturbations~\cite{intriguingSzegedy}. Instead of introducing human-imperceptible perturbations that will make a machine learning model misclassify an instance, the authors of~\cite{unrecognizableImages} efficiently produce images that are completely unrecognizable to humans, but that DNNs believe to be recognizable images with 99\% confidence. Such unrecognizable images are constructed by using Evolutionary Algorithms (EAs)~\cite{evolutionaryAlgs} or gradient ascent.

EAs are optimization algorithms inspired by Darwinian evolution. In biological terms, EAs consist of a population of “organisms” that alternately face selection in terms of keeping the best. New organisms are produced by the mutation and crossover amongst best organisms. In this case, the organisms are images and the perturbations, added to them step by step to reach the target, represent the process of mutation and crossover. The selected organisms depend on the fitness function, which, in the case of creation of images that are able to fool a machine learning classifier, is the highest confidence score the target ML model gives to that image for belonging to a specific class. 
Specifically, Nguyen  et  al.~\cite{unrecognizableImages}  use two types of encoding to produce the adversarial images, called: \textit{direct}~\cite{Stanley2003ATF} and \textit{indirect}~\cite{indirectEncoding} encoding.

\textit{EAs with direct encoding}: Each pixel value is initialized with uniform random noise within the [0, 255] range. 
Those pixel values are independently mutated. The amount of pixel values to be mutated starts at 10\% and halves after every 1000 generations. The pixel values chosen to be mutated are then altered via the polynomial mutation operator~\cite{multiObjectiveEAs} with a fixed mutation strength of 15.

\textit{EAs with indirect encoding}:
 By using indirect encoding, elements in the genome can affect multiple parts of the image~\cite{Stanley2007CompositionalPP}. Specifically, the indirect encoding used by Ngyen et al.~\cite{unrecognizableImages} is a Compositional Pattern-Producing Network (CPPN), which can evolve complex, regular images that resemble natural and man-made objects~\cite{Stanley2007CompositionalPP,Auerbach}. Unrecognizable images generated using indirect encoding are more regular as opposed to the ones produced with direct encoding, which look a lot like random noise. 

Another method used to generate unrecognizable images is by performing \textit{gradient ascent}, which involves updating an image according to the information taken by computing the gradients of the neural network the image is intended to fool.

In this paper, we modify the above-mentioned image generation approaches in order to create images that simultaneously fool a \textit{multitude} of diverse and high quality DNNs trained to solve a complex image recognition task such as ImageNet~\cite{imagenet_cvpr09} which consists of images belonging to 1,000 different classes. By creating images that fool a wide array of high quality ML models, we believe that the fooling image is able to fool even unseen classifiers that are trained to solve a similar task as the ones used to generate the image. We show that our method allows us to create unrecognizable images that are able to transfer, therefore to deceive, even unseen classifiers, thus being a promising approach to build a stronger CAPTCHA challenge.

\subsection{Adversarial Patch}
Adversarial Patch~\cite{adversarialPatch}~is a work closely related to~\cite{unrecognizableImages} and presents another method to fool DNN classifiers.
Similar to~\cite{unrecognizableImages}, the authors do not limit themselves into producing human-imperceptible perturbations like adversarial perturbations commonly do. Instead, the Adversarial Patch approach generates an image-independent patch that is extremely salient to a neural network.
This patch can then be placed anywhere within the field of view of the classifier, and causes the classifier to output the target class. The generated patch is scene-independent, and allows the authors to deceive the ML image classifiers  without prior knowledge of the lighting conditions, camera angle, type of classifier being attacked, or even the other items within the scene.

The adversarial patch solution works by completely replacing a part of the image with a generated patch.
An input patch is masked so that it takes any shape, and then it is modified via gradient descent over a variety of images, applying random translation, scaling, and rotation. The generated patch after these operations will be used as an adversarial patch to trick the ML classifiers.
In particular, for a given patch p, image x, patch location l, and patch transformations t (e.g. rotations or scaling) the authors define a patch application operator \textit{A(p, x, l, t)} which first applies the
transformations t to the patch p, and then applies the transformed patch p to the image x at location l. 
To obtain the trained patch $\hat{p}$ they use a variant of the Expectation over Transformation (EOT) framework of Athalye et al.~\cite{Athalye2018SynthesizingRA}. In particular, the patch is trained to optimize the objective function:

\[\hat{p} = \argmax\limits_p E_{ x\sim X, t\sim T, l\sim L} [log Pr(\hat{y}|A(p,x,l,t))] \]
where X is a training set of images, T is a distribution over transformations of the patch, and L is a distribution over locations in the image. This expectation is over images, thus encouraging the trained patch to work regardless of what is in the background.
The Adversarial Patch solution exploits the way image classification tasks are constructed. While images may contain several items, only one target label is considered true, and thus the network must learn to detect the most "salient" item in the frame.

The ability of the adversarial patch to impact the prediction a classifier gives to an image containing it, is highly related to the portion of the image the patch occupies. Considering that, we use adversarial patches of various sizes in our CAPTCHA scheme in order to keep both the human usability and CAPTCHA security high. As we show in the coming sections, the use of adversarial patches to strengthen image-selection CAPTCHAs is one of the strongest techniques in fooling machine learning based CAPTCHA solving bots, and also one of the most cost-efficient for the organization that uses the CAPTCHA against bots.

\section{CAPTURE: a novel CAPTCHA scheme}
\label{sec:proposal}
Our novel CAPTURE solution is based on unconstrained adversarial perturbation techniques. These techniques allow us to build CAPTCHA schemes that are able to fool ML-based bots without affecting humans ability to solve the challenge.
We identify two similar techniques to generate unconstrained adversarial examples and show how they can be used individually to build a strong CAPTCHA. We further present a more complex approach that allows us to incorporate these techniques together. 
Our proposed CAPTURE technique heavily increases the number of security features that a bot must evade in order to solve the CAPTCHA challenge, increasing the cost required to achieve a successful attack, while keeping the same level of effort for the human users.

\subsection{Threat Model}
In  our  threat  model,  the  adversary employs an automatic solvers that is incorporating machine learning classifiers to break image-selection CAPTCHA schemes. The adversary has no knowledge (black-box) on the CAPTCHA challenge generation procedure. 

The malicious Bot may employ several high quality ML-based classifiers that are trained to distinguish among the categories of the images which are part of the CAPTCHA challenge. 

The adversary is presented with a set of images in the CAPTCHA challenge and, according to the answers its classifiers gives, it decides whether to click on one or more specific images which it believes to be the answer to the challenge.

Finally, we assume that the adversary has extra tools to understand the textual description of the CAPTCHA challenge that a human should respond in order to pass the test.

\subsection{Unrecognizable Image CAPTCHA}
\label{sec:unrec_images}
The first adversarial perturbation technique we apply is unrecognizable images~\cite{unrecognizableImages}. Unrecognizable images are images that can lead machine learning classifiers into recognizing actual objects, such as a flag or curtain, when in fact the image itself only contains some peculiar visual patterns that do not resemble any real-life object. 
There are three main methods that can be used to construct unrecognizable images~\cite{unrecognizableImages}: Evolutionary Algorithms (EAs) with \textit{direct} encoding, Evolutionary Algorithms with \textit{indirect} encoding, and gradient ascent. In the following, we attempt to generate images that can fool with high confidence \textit{multiple} high quality DNNs, instead of generating images tailored against \textit{one} specific DNN classifier.

\textit{\textbf{EAs with direct encoding}}: Unfortunately, our results showed that EAs with direct encoding cannot successfully create unrecognizable images for every target class considered. Moreover, the instances that we created using EAs with direct encoding were vulnerable to simple image processing techniques, such as resizing and blurring. Due to these results, we conclude that unrecognizable images generated via direct encoding are not good candidates to build a strong CAPTCHA scheme.

\begin{figure}[!t]
\centering
\includegraphics[width=3.5in]{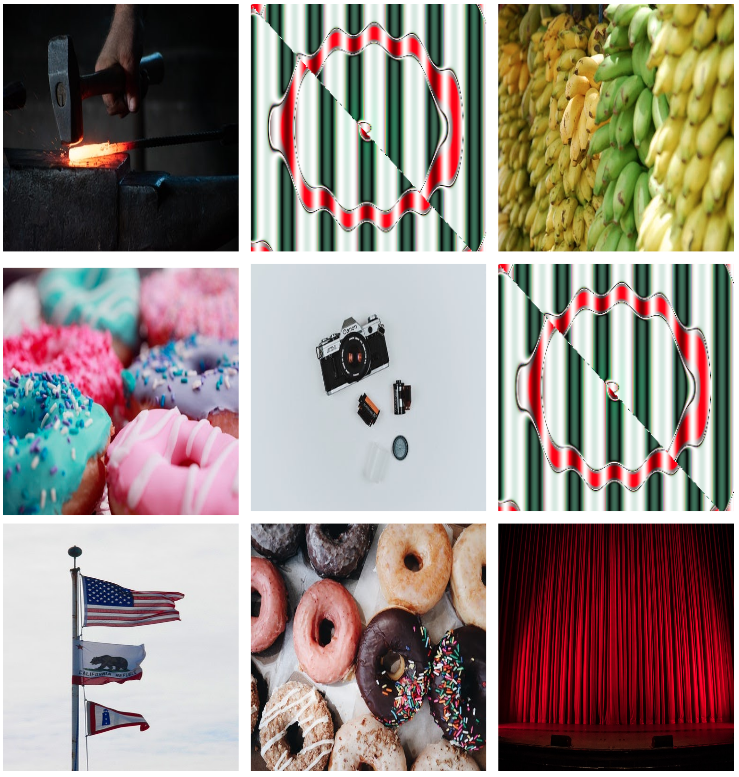}
\caption{The challenge presented to the user is: \textit{Select all the choices that show a real image of THEATER CURTAINS.} CAPTCHA with indirectly-encoded unrecognizable images.}
\label{unrecognizable_1}
\end{figure}

\textit{\textbf{EAs with indirect encoding}}: By using Compositional Pattern Producing Network (CPPN)~\cite{Stanley2007CompositionalPP} encodings, it is possible to generate unrecognizable images that fool many neural networks with high confidence. A CAPTCHA instance using unrecognizable images generated with indirect encoding is shown on Figure~\ref{unrecognizable_1}. In the CAPTCHA challenge illustrated in Figure~\ref{unrecognizable_1}, the task is to select all the images showing a theatre curtain. Out of 9 images there is only one image of a real theatre curtain (third row-third column image from the left). Among the images, we added two unrecognizable images produced via EAs with indirect encoding (first row-second image and second row-third image from the left) that successfully fool multiple, high-quality models trained on the ImageNet~\cite{imagenet_cvpr09} dataset.

While this approach is effective in generating unrecognizable images that can fool many different classifiers, it is also very computationally intensive, and the computational cost of the task vary widely based on the specific target class to which you want to assign the unrecognizable image.
Nevertheless, such image generation process can be performed offline. The unrecognizable images can be pre-computed and afterwards positioned to create the CAPTCHA challenges when needed.

Note that, unlike directly encoded unrecognizable images, indirectly encoded images are resilient to common image processing techniques.
Moreover, as an extra security feature for the CAPTCHA scheme, we can use adversarial perturbation on the clean images~\cite{adversarialCaptcha} to make it harder for the bot to distinguish between the unrecognizable images and the actual clean images. Indeed, a smart bot could use two classifiers: one to distinguish between unrecognizable and clean images, and then a second one to identify the images required by the CAPTCHA challenge among the clean images only, effectively bypassing our scheme. By applying adversarial perturbation to the clean images we can avoid this attack.

\textit{\textbf{Gradient Ascent}}: We modify the gradient ascent approach presented in~\cite{unrecognizableImages}, to create images that are able to fool an ensemble of models instead of only one. We successfully crafted unrecognizable images that look like random noise and appear meaningless to humans, but  are classified with high confidence in a predefined target class by a neural network. 
Unrecognizable images generated with gradient ascent are extremely effective, fooling most classifiers with a confidence above $99\%$. However, these type of images are not easily transferable to neural networks that were not part of the generation procedure.
Moreover, they are extremely sensitive to simple image transformations such as resizing. For example, resizing an unrecognizable image from $224\times224$ to $299\times299$ does not preserve the properties required to fool the classifier. 
Considering that the CAPTCHA is displayed in different sizes when viewed with different devices, this heavily reduces the practical applicability of the \textit{Gradient Ascent} approach. 

\subsection{Adversarially Patched CAPTCHA}
This section  studies the applicability of adversarial patches~\cite{adversarialPatch} to create a CAPTCHA challenge. This approach results to be more computationally efficient than EA with indirect encoding which is the most effective technique that we have identified in the previous section.

Adversarial patches are small patches that can be added in a region of a clean image, triggering the ML model to misclassify the image into a different specific class.
Adversarial patches can be generated efficiently and cover only a small portion of the original image, making it an appealing approach to design a resilient CAPTCHA challenge. Moreover, images modified with adversarial patches are still very easily recognized by humans, while DNNs missclassify them with high confidence.
To further strengthen our adversarial patch-based CAPTCHA scheme, we specifically design adversarial patches that fool not only a single classifier, but an ensemble of diverse, high quality classifiers such as VGG16 \& VGG19~\cite{Simonyan14verydeep}, InceptionV3~\cite{Szegedy2016RethinkingTI}, Xception~\cite{Chollet2017XceptionDL}, Resnet50~\cite{He2016DeepRL} and MobileNet~\cite{Howard2017MobileNetsEC}). Targeting an ensemble of deep neural models, rather than a specific model, greatly increases the chances that the adversarial patch produced will fool other classifiers that might be used by the CAPTCHA-solving bot.

\begin{figure}[!t]
\centering
\includegraphics[width=3.5in]{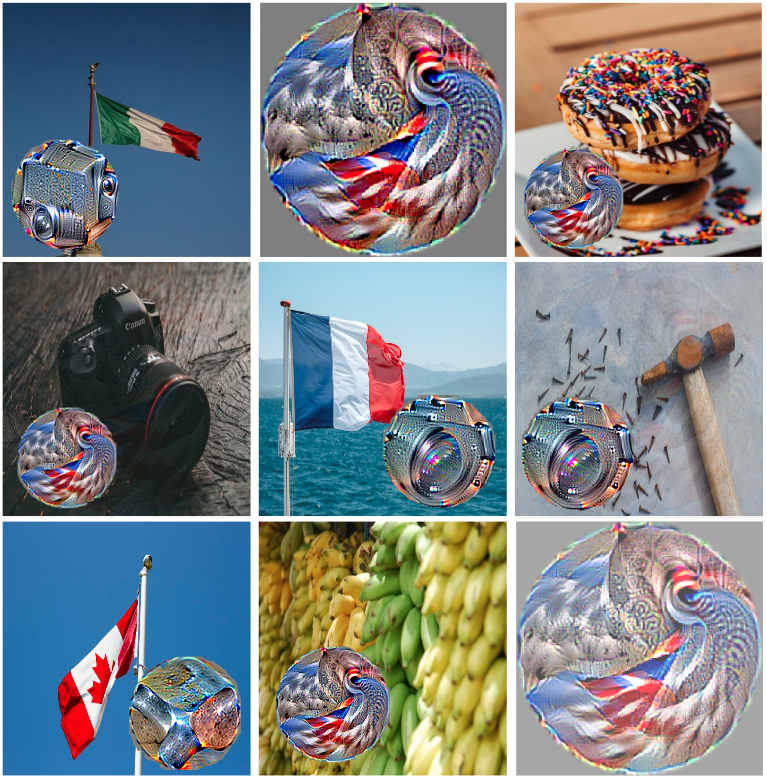}
\caption{Example of an adversarial patched CAPTCHA. The challenge presented to the user is: \textit{Select all the choices that show an image of a FLAGPOLE}.} 
\label{adv_patch_1}
\end{figure}
An instance of the CAPTCHA scheme with adversarial patches is illustrated in Figure~\ref{adv_patch_1}. The images in Figure~\ref{adv_patch_1} show various adversarial patches superimposed on pictures of several objects, such as a flagpole, a hammer and a camera. Users are asked to select all the images that include a flagpole. As we can see from the example in Figure~\ref{adv_patch_1}, this task is fairly straightforward for a human, as the adversarial patches do not cover any important feature of the original images. However, a state-of-the-art DNN would missclassify all these images with high confidence, and would be unable to identify images containing the actual flagpoles.
In this example, on the actual images of the flagpole, we superimpose patches that the model classifies as a camera or a hammer, while on all other images we superimpose patches representing a flagpole, thus increasing the chance of the bot to fail in the solution of the challenge presented.

\begin{figure}[!t]
\centering
\includegraphics[width=3.5in]{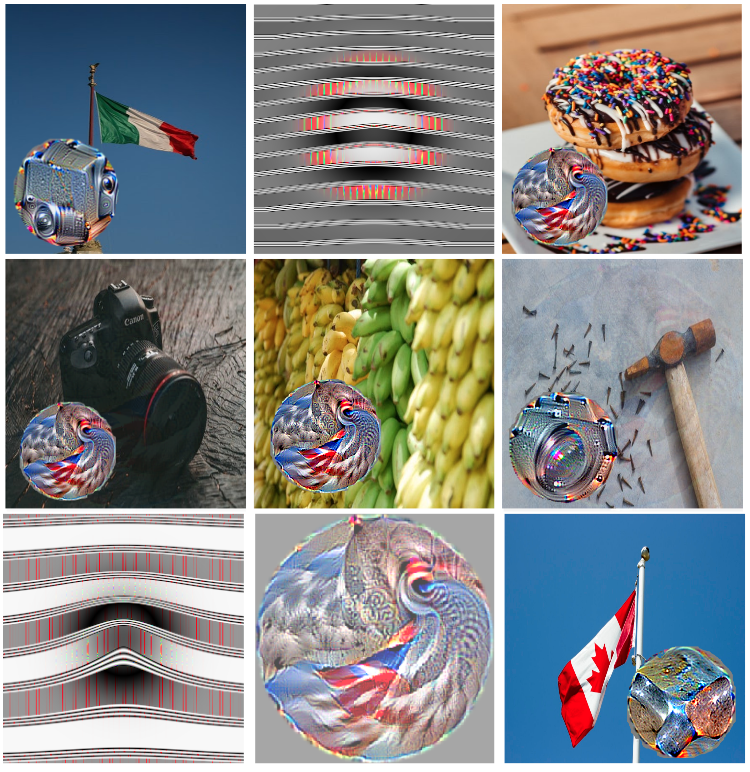}
\caption{Example of combined unrecognizable images and adversarial patching. The challenge presented to the user is: \textit{Select all the choices that show an image of a FLAGPOLE}.}
\label{adversarial_roid_1}
\end{figure}

We further integrate unrecognizable images and adversarial patches in a single CAPTCHA challenge, as illustrated in Figure~\ref{adversarial_roid_1}. The resulting scheme in far more robust
against automated ML-based bots than any single individual approach because the bot has to simultaneously evade multiple security features to solve the challenge. Moreover, combining adversarial patches and unrecognizable images provides better user experience compared to each individual scheme, as illustrated in our usability evaluation in Table~\ref{table:usability_statistics}.

\section{Experiments}
\label{sec:experiments}
A good CAPTCHA scheme is one that is secure while not compromising the human ability to solve it. The following sections present our experiments both on security and on user experience aspects of our proposed CAPTURE scheme.

\subsection{Security Evaluation}

Our security evaluation models the image-selection CAPTCHA as a challenge where images belong to the categories of the ImageNet~\cite{imagenet_cvpr09} image recognition dataset. 
Malicious ML-based Bots could use state-of-the-art ML models to identify the CAPTCHA image categories thus solving the challenge.
Our evaluation assumes that the malicious Bots utilize the best performing ML models for image recognition available today, such as VGG16 \& VGG19~\cite{Simonyan14verydeep}, ResNet50~\cite{He2016DeepRL}, InceptionV3~\cite{Szegedy2016RethinkingTI}, Xception~\cite{Chollet2017XceptionDL} and MobileNets~\cite{Howard2017MobileNetsEC}.
All these ML models have a published accuracy of above 80\% in correctly classifying an image to its correct category. Thus, a CAPTCHA challenge based on real images (e.g. without the presence of unrecognizable images or adversarial patches) is vulnerable, and can be solved with 80\% accuracy by Bots utilizing the above ML models.
Our experiments show that introducing the proposed security features, i.e. unrecognizable images and adversarial patch, we can decrease the accuracy of these advanced ML models almost to zero percent, and significantly reduce the ability of the ML-based Bots to bypass our CAPTURE challenge.

\begin{itemize}
    \item \textit{Unrecognizable Images}: We generate unrecognizable images with indirect encoding that are able to fool an ensemble of diverse and high quality DNNs. The neural networks employed on our experiments are: VGG16 \& VGG19~\cite{Simonyan14verydeep}, ResNet50~\cite{He2016DeepRL}, InceptionV3~\cite{Szegedy2016RethinkingTI}, Xception~\cite{Chollet2017XceptionDL} and MobileNets~\cite{Howard2017MobileNetsEC}. We generate unrecognizable images that are able to deceive 5 of the above neural networks, and we also assess whether that ability to deceive is transferred to the sixth held out DNN. We generate 1200 unrecognizable images (200 per each ensemble combination). On average the unrecognizable images were able to deceive the held-out DNN around 70\% of the time with over 95\% confidence in the prediction.
    
    The achieved fooling rates on the unrecognizable images highlight that the image generation procedure needs to be improved to achieve higher transferability rates. 
    
    \item \textit{Adversarial Patch}: in the case of adversarial patches we used for the experiments the same set of high quality DNNs as in the case of unrecognizable images. We generate 1200 adversarial patches (200 per each ensemble combination). Then, we evaluate their fooling capability by randomly putting them on 500 different images starting from 10\% up to 100\% of the image size and evaluate the ability of the patch to fool the remaining neural network to check how the generated adversarial patch transfers to that ML model. Generating an adversarial patch that is able to simultaneously fool multiple diverse and high quality ML models makes the patch capabilities more general and possibly effective even on other unseen ML models, like the one that the malicious Bot might be using.

Figure~\ref{adversarial_patch_success} displays the success rate on each of the unseen models. 
\begin{figure}[!h]
\centering
\includegraphics[width=3.5in]{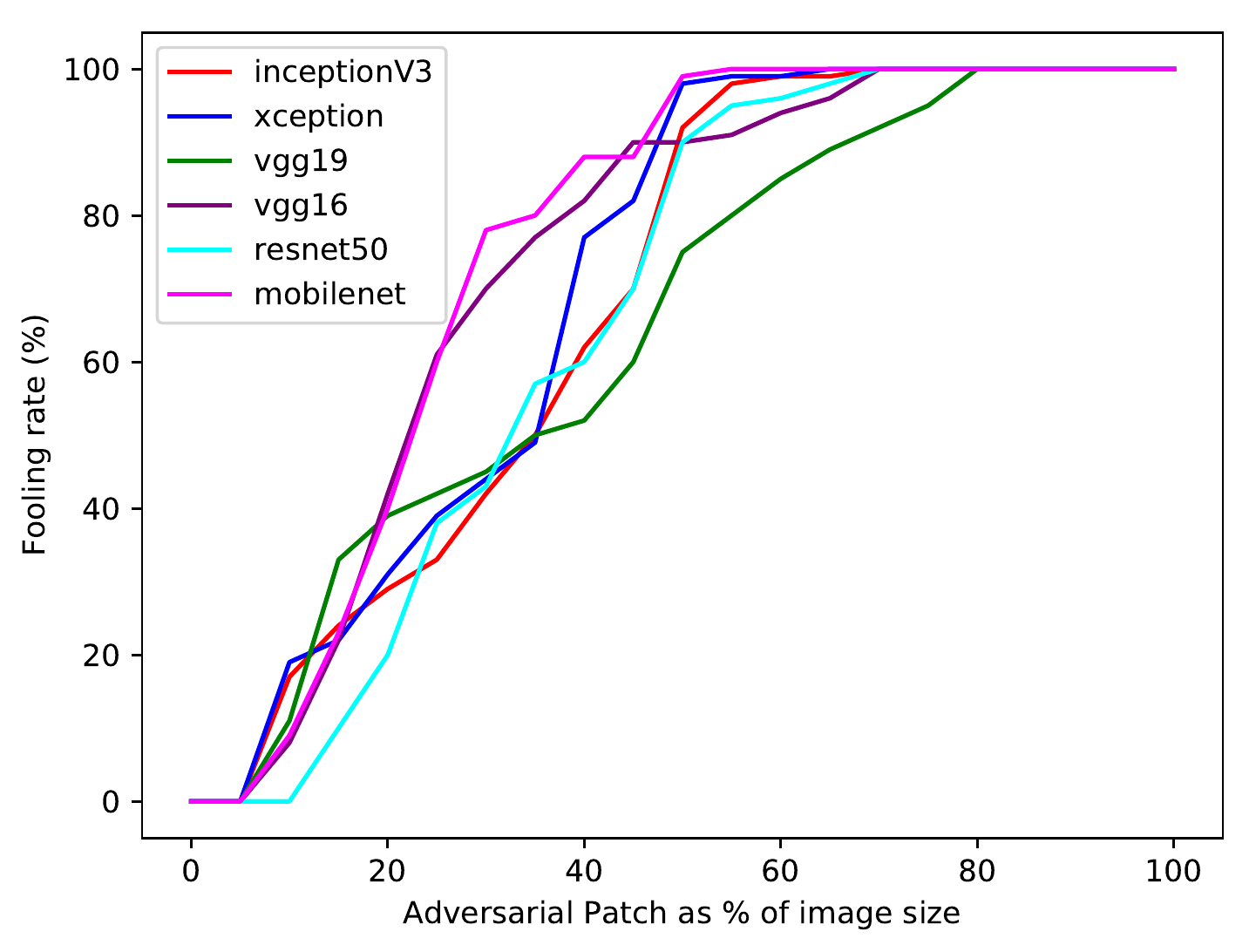}
\caption{Adversarial Patch success rates. The lines represent the ability of the adversarial patch of different size (generated on the white-box access models), to fool the black-box model. In the x-axis, the percentage 100\% means that the patch is as large as the whole image. With 50\% the patch is one quarter of the entire image}
\label{adversarial_patch_success}
\end{figure}
Figure~\ref{adversarial_patch_success} shows that an adversarial patch scaled to a factor close to 60\% is able to fool high quality classifiers almost 100\% of the time. In addition, having a patch that covers slightly more than a quarter of the image, provides a better user experience in the solving process also. Since the patch covers a small portion of the image, we can make the object(s) that the user should choose to solve the CAPTURE challenge, simpler to identify to human users.
\end{itemize}

\subsection{Usability Evaluation}
We tested our various configurations of CAPTURE challenge with a group of users of different age groups. Each person was requested to solve 10 CAPTURE challenges similar to the examples displayed on Figures~\ref{unrecognizable_1},~\ref{adv_patch_1}, and~\ref{adversarial_roid_1}. In total, we surveyed 113 people. Table~\ref{table:user_stats} shows the number of participants per age group. In Table~\ref{table:usability_statistics}, we report the success rate of the participants in solving the CAPTCHA challenges presented. Note that almost all users have been able to successfully solve the CAPTCHA challenges with a success rate of above 85\%. 

\begin{table}[!h]
\centering
\begin{tabular}{|c|c|c|c|c|c|c|}
\hline
Age & [16-20] & [21-30] & [31-40] & [41-50] & [51-60] & [61+]\\ \hline
Surveyed & 1 & 72 & 5 & 12 & 19 & 4\\ \hline
\end{tabular}

\caption{User Statistics}
\label{table:user_stats}
\end{table}

\begin{table}[!h]
\centering
\begin{tabular}{|c|c|c|c|}
\hline
  Challenge & Unrec. Images & Adv. Patch & Unrec. + Adv. Patch \\\hline
  Success Rate & 96.5\% & 86\% & 92.3\% \\\hline
\end{tabular}

\caption{Usability Statistics. (The reported success rate is the weighted average with respect to the percentage of participants per age group.)}
\label{table:usability_statistics}
\end{table}

When asked to compare their experience with other image selection CAPTCHAs, the users said our CAPTURE task was easier to solve because the images the user should click were easy to find. The presence of weird-looking images did not prevent them from solving the task, and most of them said that it was even easier because they could discard whole unrecognizable images, and thus answer the challenge faster.

After carefully investigating the survey results and carrying out post interviews with the surveyed persons, we found out that some mistakes were accountant to people not paying attention to the challenge description.
However, this does not impede the applicability of our scheme, as in real-life applications, if human users make a mistake, they are typically presented with another CAPTCHA challenge in order to access the required web-service.

\section{Conclusions}
\label{sec:conclusions}

This paper presents a new enhancement of the CAPTCHA challenge called: CAPTURE - CAPtcha Technique Uniquely REsistant. The new CAPTURE scheme is based on a "good use" of the recent advancements in adversarial machine learning.
In particular, this paper proposes a novel CAPTCHA scheme that employs two approaches to create human-unrecognizable images that are recognizable by machine learning models as a real object with high confidence. 
The core advantages compared to traditional image CAPTCHAs include:
\begin{itemize}
    \item \textit{ML-based Bots cannot easily solve the CAPTURE challenge}: Using unrecognizable images and adversarial patches,
    the CAPTURE scheme can fool the Bot into clicking on images that the human user would never click to solve the challenge. Also, incorporating multiple fooling strategies like unrecognizable images~\cite{unrecognizableImages}, adversarial patches~\cite{adversarialPatch}, and adversarial perturbations~\cite{intriguingSzegedy} in the same CAPTCHA, we make it harder for the Bot to bypass all the security measures in each of the images that are part of the CAPTURE challenge presented to it.
    \item \textit{Improve the user experience while solving the CAPTCHA challenge}: Using unrecognizable images and adversarial patches throughout the images that the user should not click to solve the challenge (but the bots will be prone to click) allows us to put clear images of the objects the user should click thus making their life easier. Current image CAPTCHAs employ images that are hard to distinguish even to human users which negatively impact the user experience. Our CAPTURE scheme increases the security while also keeping a positive level in terms of user experience in the solving process.
\end{itemize}

As future work, we intend to make the generation of unrecognizable image more computationally feasible. We also intend to focus on increasing the ability of our mechanism to mislead multiple ML models simultaneously.
Based on our users feedback, this would lead to better-looking CAPTCHAs for humans, while also increasing the robustness against ML-based Bot attacks.

\section{Acknowledgments}
The work of Dorjan Hitaj and Luigi V. Mancini was supported by Gen4olive, a project that has received funding from the European Union’s Horizon 2020 research and innovation programme under grant agreement No. 101000427, and in part by the Italian MIUR through the Dipartimento di Informatica, Sapienza University of Rome, under Grant Dipartimenti di eccellenza 2018–2022. The work of Sushil Jajodia was supported by the Office of Naval Research grant N00014-18-1-2670 and by the Army Research Office grant W911NF-13-1-0421.

\bibliographystyle{unsrt}
\bibliography{bibliography.bib}

\begin{thebibliography}{10}

\bibitem{unrecognizableImages}
A.~{Nguyen}, J.~{Yosinski}, and J.~{Clune}.
\newblock Deep neural networks are easily fooled: High confidence predictions
  for unrecognizable images.
\newblock In {\em 2015 IEEE Conference on Computer Vision and Pattern
  Recognition (CVPR)}, pages 427--436, 2015.

\bibitem{adversarialPatch}
Tom~B. Brown, Dandelion Man{\'{e}}, Aurko Roy, Mart{\'{\i}}n Abadi, and Justin
  Gilmer.
\newblock Adversarial patch.
\newblock {\em CoRR}, abs/1712.09665, 2017.

\bibitem{intriguingSzegedy}
Christian Szegedy, Wojciech Zaremba, Ilya Sutskever, Joan Bruna, Dumitru Erhan,
  Ian~J. Goodfellow, and Rob Fergus.
\newblock Intriguing properties of neural networks.
\newblock In {\em 2nd International Conference on Learning Representations,
  {ICLR} 2014, Banff, AB, Canada, April 14-16, 2014, Conference Track
  Proceedings}, 2014.

\bibitem{papernot2016transferability}
Nicolas Papernot, Patrick McDaniel, and Ian Goodfellow.
\newblock Transferability in machine learning: from phenomena to black-box
  attacks using adversarial samples.
\newblock {\em arXiv preprint arXiv:1605.07277}, 2016.

\bibitem{ensembleTramer}
Florian Tram{\`e}r, Alexey Kurakin, Nicolas Papernot, Dan Boneh, and Patrick~D.
  McDaniel.
\newblock Ensemble adversarial training: Attacks and defenses.
\newblock {\em CoRR}, abs/1705.07204, 2017.

\bibitem{adversarialCaptcha}
Chenghui Shi, Xiaogang Xu, Shouling Ji, Kai Bu, Jianhai Chen, Raheem~A. Beyah,
  and Ting Wang.
\newblock Adversarial captchas.
\newblock {\em CoRR}, abs/1901.01107, 2019.

\bibitem{evolutionaryAlgs}
Dario Floreano and Claudio Mattiussi.
\newblock {\em Bio-Inspired Artificial Intelligence: Theories, Methods, and
  Technologies}.
\newblock The MIT Press, 2008.

\bibitem{Stanley2003ATF}
Kenneth~O. Stanley and Risto Miikkulainen.
\newblock A taxonomy for artificial embryogeny.
\newblock {\em Artificial Life}, 9:93--130, 2003.

\bibitem{indirectEncoding}
Jeff Clune, Kenneth~O. Stanley, Robert~T. Pennock, and Charles Ofria.
\newblock On the performance of indirect encoding across the continuum of
  regularity.
\newblock {\em IEEE Transactions on Evolutionary Computation}, 15:346--367,
  2011.

\bibitem{multiObjectiveEAs}
Kalyanmoy Deb.
\newblock {\em Multi-Objective Optimization Using Evolutionary Algorithms}.
\newblock John Wiley \& Sons, Inc., USA, 2001.

\bibitem{Stanley2007CompositionalPP}
Kenneth~O. Stanley.
\newblock Compositional pattern producing networks: A novel abstraction of
  development.
\newblock {\em Genetic Programming and Evolvable Machines}, 8:131--162, 2007.

\bibitem{Auerbach}
Joshua~E. Auerbach.
\newblock Automated evolution of interesting images.
\newblock {\em Artificial Life 13}, 2012.
\newblock The Humanities and ALife -- Best Presentation Award Winner.

\bibitem{imagenet_cvpr09}
J.~Deng, W.~Dong, R.~Socher, L.-J. Li, K.~Li, and L.~Fei-Fei.
\newblock {ImageNet: A Large-Scale Hierarchical Image Database}.
\newblock In {\em CVPR09}, 2009.

\bibitem{Athalye2018SynthesizingRA}
Anish Athalye, Logan Engstrom, Andrew Ilyas, and Kevin Kwok.
\newblock Synthesizing robust adversarial examples.
\newblock In {\em ICML}, 2018.

\bibitem{Simonyan14verydeep}
Karen Simonyan and Andrew Zisserman.
\newblock Very deep convolutional networks for large-scale image recognition,
  2014.

\bibitem{Szegedy2016RethinkingTI}
Christian Szegedy, Vincent Vanhoucke, Sergey Ioffe, Jonathon Shlens, and
  Zbigniew Wojna.
\newblock Rethinking the inception architecture for computer vision.
\newblock {\em 2016 IEEE Conference on Computer Vision and Pattern Recognition
  (CVPR)}, pages 2818--2826, 2016.

\bibitem{Chollet2017XceptionDL}
François Chollet.
\newblock Xception: Deep learning with depthwise separable convolutions.
\newblock {\em 2017 IEEE Conference on Computer Vision and Pattern Recognition
  (CVPR)}, pages 1800--1807, 2017.

\bibitem{He2016DeepRL}
Kaiming He, Xiangyu Zhang, Shaoqing Ren, and Jian Sun.
\newblock Deep residual learning for image recognition.
\newblock {\em 2016 IEEE Conference on Computer Vision and Pattern Recognition
  (CVPR)}, pages 770--778, 2016.

\bibitem{Howard2017MobileNetsEC}
Andrew~G. Howard, Menglong Zhu, Bo~Chen, Dmitry Kalenichenko, Weijun Wang,
  Tobias Weyand, Marco Andreetto, and Hartwig Adam.
\newblock Mobilenets: Efficient convolutional neural networks for mobile vision
  applications.
\newblock {\em CoRR}, abs/1704.04861, 2017.

\end{thebibliography}

\end{document}